\documentclass[%
 reprint,
superscriptaddress,
 amsmath,amssymb,
 aps,linenumbers,
]{revtex4-2}

\usepackage{xcolor}
\usepackage{graphicx}
\usepackage{dcolumn}
\usepackage{bm}
\usepackage{hyperref}
\usepackage{amsmath,esint,eqnarray}
\usepackage{verbatim}

\newcommand{\beginsupplement}{%
        \setcounter{table}{0}
        \renewcommand{\thetable}{S\arabic{table}}%
        \setcounter{figure}{0}
        \renewcommand{\thefigure}{S\arabic{figure}}%
     }

\begin{document}

\nolinenumbers

\title{Probing active nematics with in-situ microfabricated elastic inclusions}

\author{Ignasi V\'elez-Cer\'on}
\affiliation{Department of Materials Science and Physical Chemistry, Universitat de Barcelona, 08028 Barcelona, Spain}
\affiliation{Institute of Nanoscience and Nanotechnology, IN2UB, Universitat de Barcelona, 08028 Barcelona, Spain}
\author{Pau Guillamat}%
\affiliation{%
 Institute for Bioengineering of Catalonia (IBEC), The Barcelona Institute for Science and Technology (BIST), Barcelona, 08028, Spain}%
\author{Francesc Sagu\'es}
\affiliation{Department of Materials Science and Physical Chemistry, Universitat de Barcelona, 08028 Barcelona, Spain}
\affiliation{Institute of Nanoscience and Nanotechnology, IN2UB, Universitat de Barcelona, 08028 Barcelona, Spain}

\author{Jordi Ign\'es-Mullol}
\affiliation{Department of Materials Science and Physical Chemistry, Universitat de Barcelona, 08028 Barcelona, Spain}
\affiliation{Institute of Nanoscience and Nanotechnology, IN2UB, Universitat de Barcelona, 08028 Barcelona, Spain}
 \email{jignes@ub.edu}

\date{\today}

\begin{abstract}
In this work, we report a direct measurement of the forces exerted by a tubulin/kinesin active nematic gel as well as its complete rheological characterization, including the quantification of its shear viscosity, $\eta$, and its activity parameter, $\alpha$. For this, we develop a novel method that allows us to rapidly photo-polymerize compliant elastic inclusions in the continuously remodelling active system. Moreover, we quantitatively settle long-standing theoretical predictions, such as a postulated relationship encoding the intrinsic time scale of the active nematic in terms of $\eta$ and $\alpha$. In parallel, we infer a value for the nematic elasticity constant, $K$, by combining our measurements with the theorized scaling of the active length scale. On top of the microrheology capatilities, we demonstrate novel strategies for defect encapsulation, quantification of defect mechanics, and defect interactions, enabled by the versatility of the new microfabrication strategy that allows to combine elastic motifs of different shape and stiffness that are fabricated in-situ and on-time.\\

\end{abstract}

\maketitle

Active matter embraces a wide variety of systems composed of dense collections of units that continuously convert energy into forces, thus being inherently out of equilibrium \cite{sagues22}. Interactions between these force-generating units lead to collective phenomena such as flocking, seemingly-turbulent flows, and other emergent mechanodynamical patterns \cite{Ramaswamy10, Marchetti13}. Besides the many natural examples that can be found at different length scales \cite{Herbert-Read11, Bialek12, Buhl06}, similar phenomena can be obtained in-vitro by bottom up approaches based on the densification of active colloidal particles including, for instance, monolayers of mammalian cells \cite{Saw17, Kawaguchi17, Blanch-Mercader18, Duclos18}, self-propelled particles \cite{Theurkauff12, Palacci13, Buttinoni13, Wang15Accounts}, living bacteria \cite{Zhang10, Wensink12, Li19, Turiv20} or reconstitutions of cytoskeletal filaments and motor proteins \cite{Sanchez12, Martinez19}.

The latter systems are among the most fascinating examples in the field of synthetic active matter \cite{Needleman17}. At the expenses of adenosine triphosphate (ATP) consumption, bound molecular motor clusters internally shear protein filament bundles, leading to continuous chaotic flows. In a dense, quasi-two-dimensional phase, known as active nematic (AN) \cite{Doost18,Zhang21}, these active gels feature continuously changing textures with liquid-crystal-like orientational (nematic) order. The latter is lost in local regions that configure topological defects.

Experiments have been able to characterize \cite{Guillamat16PRE, Tan19, Martinez21}, even control \cite{Guillamat16, Guillamat17, Zhang21} and manipulate the dynamics of AN realisations \cite{Keber14,Wu17, Hardouin}, and theoretical and numerical works have successfully rationalized experimental observations \cite{Giomi15, Zhang16, Alert19, Pearce19, Thijssen21}. In spite of remarkable recent studies with monolayers of mammalian cells featuring nematic order, \cite{Saw17} little is known about the active mechanics of AN materials. More fundamentally, an accurate quantitative determination of the most important constitutive parameters of the microtubule-based system is lacking. Here, we address this challenge by embedding microfabricated elastic inclusions with pre-designed size and geometry within the AN layer. By studying the deformations of such objects we have been able to probe the system's rheology and mechanics.

\begin{figure*}[t]
	\centering
	\includegraphics[width=0.8\textwidth]{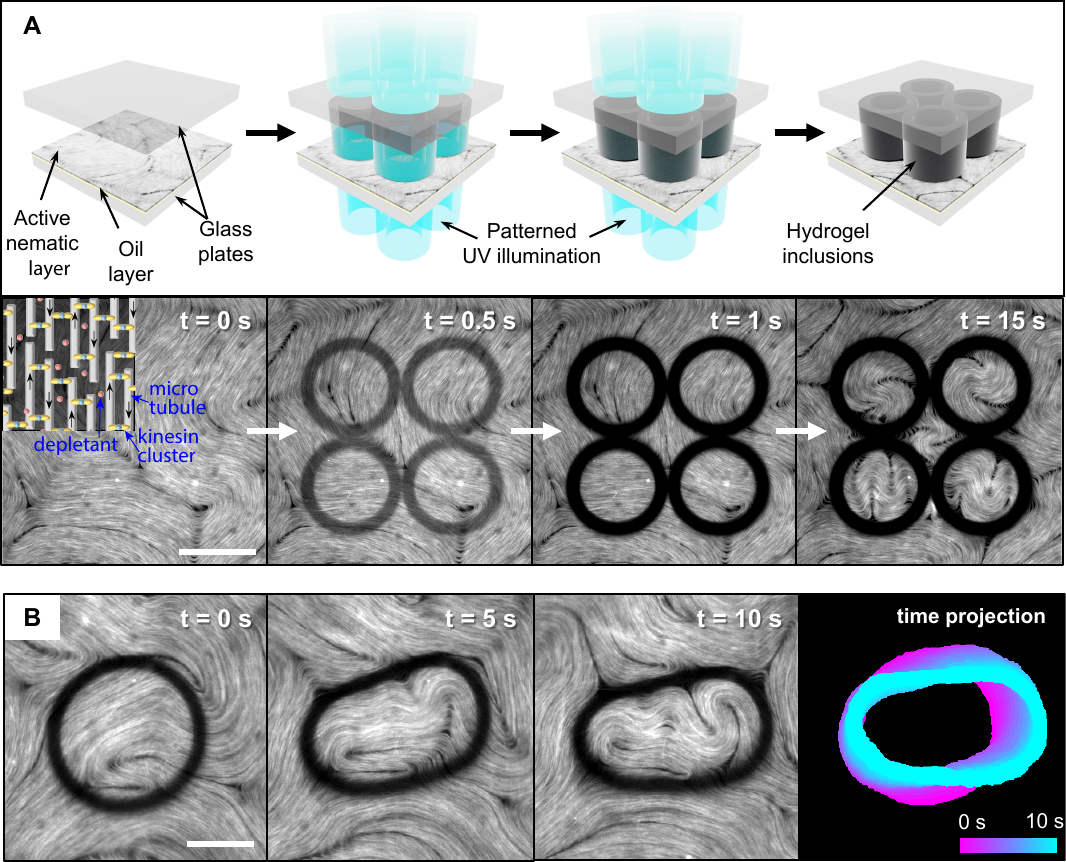}
	\caption{In situ photo-polymerization of hydrogel structures in an active nematic. (A) (Top) Sketch of the experimental cell and polymerization of four rigid hydrogel rings within the AN (See SI Video \ref{SMovie:rings} and also Fig. \ref{SFig:optics} for further details of the setup). (Bottom) Fluorescence micrographs before irradiation and during the emergence of the hydrogel structures, and the subsequent adaptation of the AN to the sudden confinement. Elapsed times are overlaid on each frame. The inset in the leftmost micrograph includes a sketch with the microstructure of the AN filaments, where antiparallel microtubules slide due to the action of kinesin dimers. The presence of a depletant agent favours the condensation of the dispersed biomaterials.
(B) Shorter irradiation times lead to the formation of a deformable columnar ring (fluorescence micrographs; see also SI Video \ref{SMovie:flexible}). Elapsed times from the first panel, where the ring has already been created, are overlaid on each frame. The rightmost panel illustrates the deformation of the inclusion over time.
Scale bars, $50\;\mu$m.
}
	\label{fig:setup}
\end{figure*}

\begin{figure*}[t]
	\centering
	\includegraphics[width=0.9\textwidth]{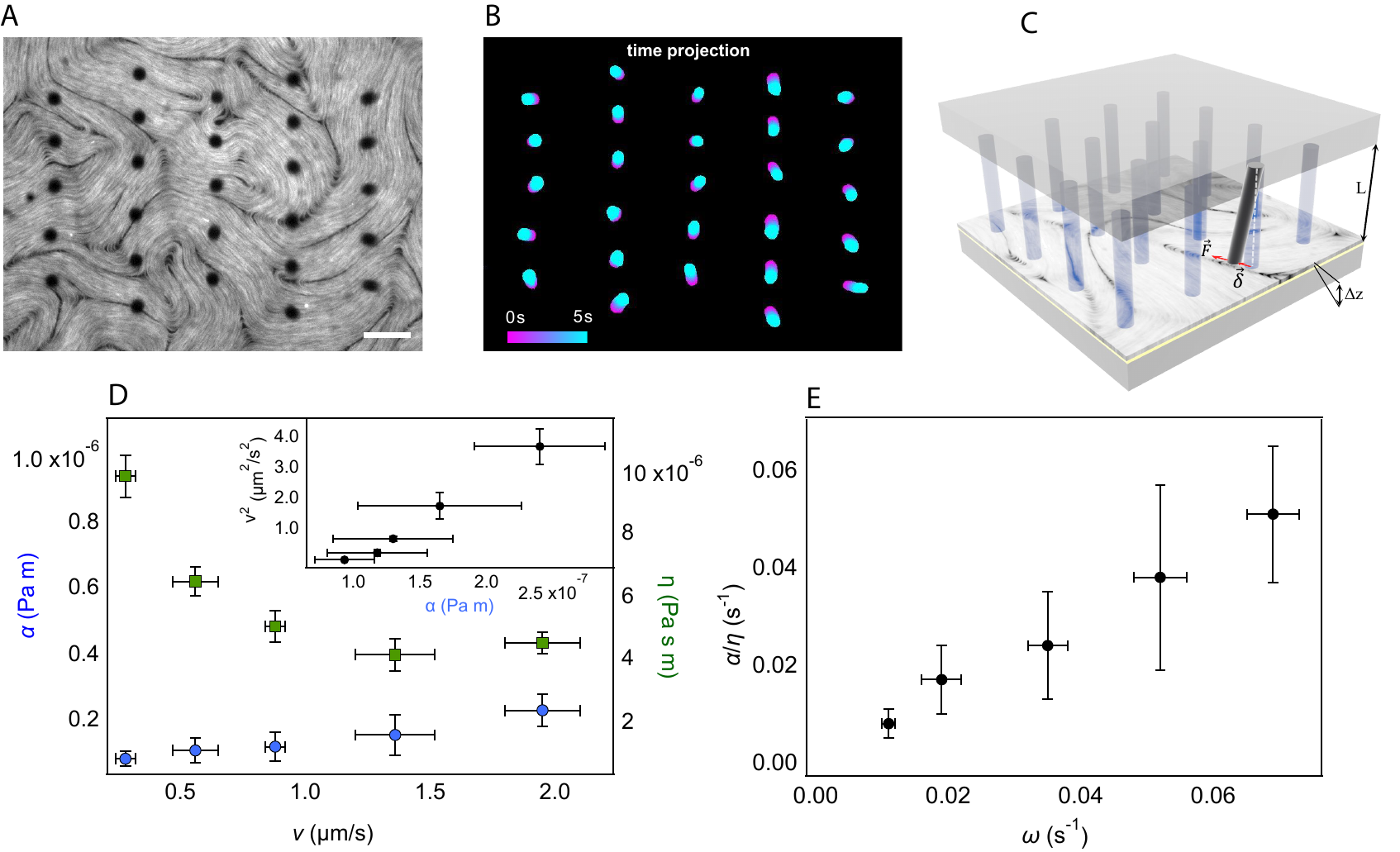}
	\caption{Probing AN forces with elastic micro-pillars. (A) Fluorescence micrograph of an AN layer at the instant when a lattice of soft columns has been photo-polymerized (see also SI Movie \ref{SMovie:lattice}). Irradiation time is 1.5 s. Scale bar is $50\;\mu$m. (B) Color-coded location of the columns' free surface during the first 5 s after polymerization. (C) The flexible columns are deflected by $\vec{\delta}$ due to the action of $\vec{F}$, the resultant of the viscous drag and active forces acting at free end of each column, which are submerged for a distance $\Delta z$ in the AN layer. (D) Measured activity parameter and viscosity of the AN layer vs. the average speed of the AN flows. In the inset, the square of the speed is plotted vs. the activity parameter. (E) Comparison between the inverse of the active time scale, $\alpha/\eta$, computed from the parameters reported in panel (D), and the average vorticity, $\omega$, obtained from the active flow velocity map. In (D) and (E), error bars are the standard deviation of the mean for $v$ and $\omega$, and confidence intervals for the fitted parameters $\alpha$ and $\eta$.
}
	\label{fig:balance}
\end{figure*}

To this end, we tune our active preparations by adding to the standard formulation of a kinesin/tubulin active gel the precursors of a hydrogel that can be photo-polymerized with UV light. This new mixture is introduced in a flow cell with a gap of 50 microns between two glass plates (see SI, Sec. \ref{sec:Methods}, for full details). One plate incorporates a superhydrophilic polyacrylamide (PAM) brush, which prevents protein adsorption. The other plate is functionalized with fluorinated alkane chains that adsorb a thin layer of a fluorinated oil. The active nematic forms at the water/oil interface, which is decorated with a polyethyleneglycol-based surfactant that prevents biomolecules from being in direct contact with the oil.
As stated above, apart from the standard components of the active preparation, the aqueous subphase incorporates polyethyleneglycol-acrylate monomers and lithium phenyl-2,4,6-trimethylbenzoylphosphinate (LAP)\cite{Fairbanks09}, a photosensitive molecule (see Table \ref{table:Conc}). Irradiation with UV light induces the dissociation of LAP into radical species that initiate a chain polymerization reaction leading to the formation of a PEG hydrogel in the regions exposed to the UV light. In our experiments, we use a custom illumination setup based on a digital micromirror device (DMD) projector integrated into the light path of an inverted fluorescence microscope (Fig. \ref{SFig:optics}). Once the AN is formed, patterns of UV light are focused on the AN layer by means of the microscope objective, which sets the maximum size and lateral resolution of the built objects, as well as the range of power densities of the irradiation.

With this setup, we can generate in-situ within the active gel columnar hydrogel microstructures of arbitrary shape, set by the illumination pattern, and stiffness, according to the irradiation time and footprint of the objects. Columns are grafted on the PAM-coated plate and extend across the aqueous phase, up to the AN layer in contact with the oil phase. In Fig. \ref{fig:setup}A we show schematically the process to create rigid hydrogel columns of arbitrary shape anywhere and at any time on an evolving AN layer. As an example, in Fig. \ref{fig:setup}A four adjacent rings are created, as seen in the fluorescence micrographs below the sketches. The AN layer adapts to the new boundary conditions by transforming the local defect pattern to accommodate the imposed topology. Notice that the fluorescence probe contained in the hydrogel volume is photobleached by the action of UV light, thus appearing dark in the fluorescence micrographs upon polymerization. In Fig. \ref{fig:setup}B, we demonstrate the system's capability of generating also soft columnar objects. In this case, a flexible ring that is continuously deformed by forces, generated by the interplay of the ANs inside and outside the inclusion. Both rigid and flexible walls trigger the formation of active boundary layers \cite{Hardouin22} that determine the geometry of the enclosed AN. The high versatility of this protocol allows to print, either in sequence or simultaneously, both rigid and soft, deformable structures. This can be demonstrated by imprinting a combination of flexible pillars enclosed by a rigid ring; see SI. Video \ref{SMovie:hybrid}. This allows both to monitor the adaptation of the AN to a sudden change in boundary conditions, and to probe the local stresses exerted by the active material, as we discuss below.

In this study, we take advantage of the elasticity of the soft hydrogel to imprint an array of compliant pillars with circular cross-section as microrheological probes for the AN layer (Fig. \ref{fig:balance}A,B). Each pillar will act as a cantilever, anchored on the hydrophilic plate and bent by the action of the AN layer. The active material exerts a combination of hydrodynamic, active, and elastic stresses that results in the so-called active turbulent regime \cite{Giomi15, Martinez21}. Under mechanical equilibrium, column bending will be the result of the net forces arising in the active fluid. Previous studies with this material suggested that hydrodynamic and active stresses have similar orders of magnitude, while elastic stresses, being of higher order in the gradients of the orientational field, are subdominant except very near topological defects \cite{Joshi22,Golden23}. We will take this into account and assume in our system that column deflection is balanced by the effect of active and hydrodynamic (shear) stresses alone.

\begin{figure*}[t]
	\centering
	\includegraphics[width=0.9\textwidth]{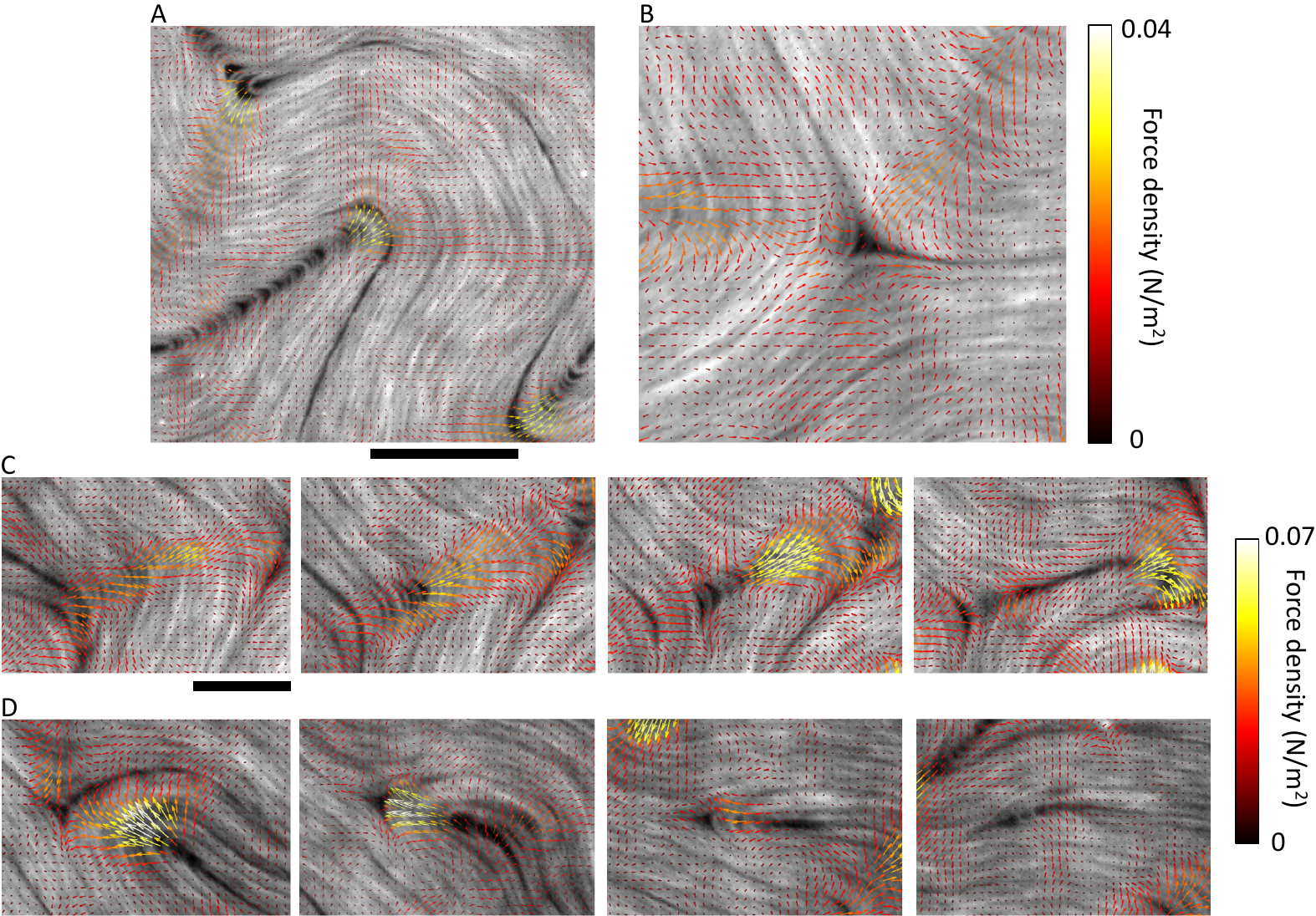}
	\caption{Active nematic forces. (A) Force density map, obtained from $f_i = 2\eta \partial_j\varepsilon_{ij} -\alpha \partial_j Q_{ij}$, in the vicinity of a $+1/2$ defect. The force density accumulates at the tip of the self-propelled defect. (B) In the vicinity of a $-1/2$ defect, the force density exhibits nearly three-fold symmetry. (C) Force density map during the unbinding of a defect pair. The force density accumulates at the tip of the bending filaments, leading to the formation of a $+1/2$ and $-1/2$ defect pair. pair. Elapsed times from the first frame are 2 s, 8 s,
and 12 s (see also SI Video \ref{SMovie:unbind}). (D) Force density map during the annihilation of a defect pair. The high force density at the tip of the moving $+1/2$ defect vanishes after coalescence. Elapsed times from the first frame are 9 s, 14 s, and 19 s (see also SI Video \ref{SMovie:bind}). Scale bars, 50 $\mu$m.
}
	\label{fig:forcemaps}
\end{figure*}

For the characterization of active forces, we balance the elastic pillar deflection force (see SI  Sec. \ref{sec:balance}) with the stresses within the active material, $\sigma^{AN}_{ij}$, computed on the thin corona of the hydrogel cylinder of thickness $\Delta z$ that is submerged in the AN layer (Fig. \ref{fig:balance}C). In the following, we also assume that the force is applied at the edge of the free surface of the pillar. Moreover, we ignore the viscous drag due to the oil layer in contact with the upper cylinder' surface and that due to the aqueous volume that embraces most of the column, both negligible (see SI  Sec. \ref{sec:drag} for details of this analysis).

The force exerted by the AN on a pillar is computed as $F_i = \oint \sum_{j=1,2} \sigma^{AN,2d}_{ij}n_j dl$, where $i,j$ span the two spatial dimensions, $n_i$ is the unit vector normal to the lateral surface of the column, the line integral spans the circular cross section of the column, and $\sigma^{AN,2d}_{ij}$ is the two-dimensional AN stress tensor (see SI Sec. \ref{sec:balance}). In terms of the dominant viscous and active stresses, $\sigma^{AN,2d}_{ij} = 2\eta \varepsilon_{ij} -\alpha Q_{ij}$, where $\varepsilon_{ij}$ and $Q_{ij}$ denote, respectively, the 2d strain rate tensor and nematic tensor order parameter. The parameters $\eta$ and $\alpha$, which we measure here, stand respectively for the 2d shear viscosity and activity coefficient of the AN (assumed positive for our extensile system \cite{Doost18}).

On the other hand, the force required for the bending deflection of the pillar is $F_i^{def}=\frac{3\delta_i E I}{L^3}$, where $\delta_i$ denotes the in-plane component of the deflection, while $E$, $I$, and $L$ correspond respectively to the Young's modulus (which we determine using fluids of known viscosity, see SI, Sec.\ref{sec:Young}), the second moment of inertia of the cylindrical column, and the height of the pillars, set by the spacing between glass plates (see Fig. \ref{fig:balance}C).

After their photopolymerization within the AN, we will assume that each pillar achieves mechanical equilibrium either when its deflection reaches a steady-state value or just before it changes direction (see SI  Sec. \ref{sec:balance}). Moreover, local shear and active stress will be computed from the instantaneous orientation and velocity fields of the AN, as defined above, just before pillar polymerization. In our analysis, we have included columns of different diameters in the range $10.4\;\mu$m-$18\;\mu$m, to achieve a deflection at mechanical equilibrium in the range $2\;-14\;\mu$m for all ATP concentrations. Our measurements show that the total force exerted on a single column is in the range 1 - 20 pN, depending on the column diameter and local fields of the AN. In all cases, the drag force due to active flows is typically 3-4 times larger than the corresponding force due to active stress.

\begin{figure*}[h]
	\centering
	\includegraphics[width=0.8\textwidth]{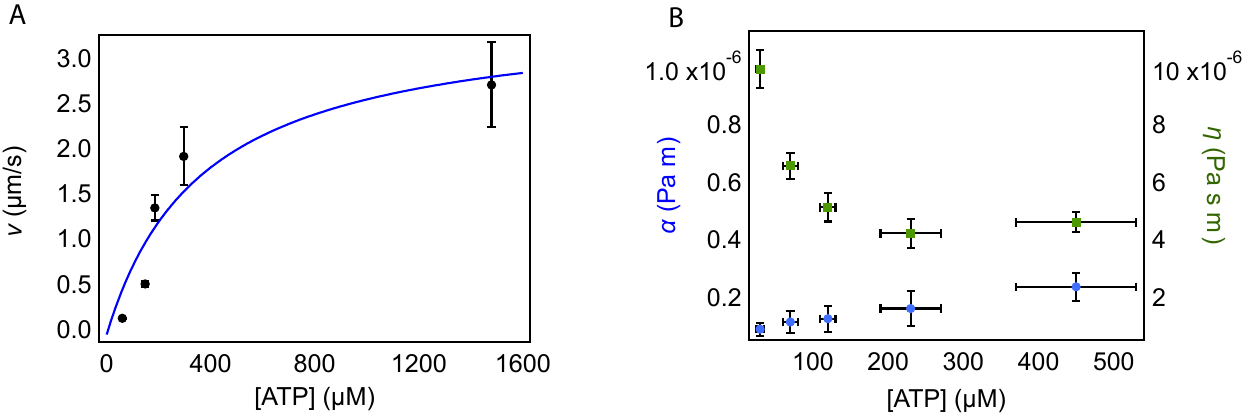}
	\caption{(A) Calibration of the average AN speed as a function of ATP concentration. The solid line is a fit to a Michaelis-Menten model with fitted parameters $v_{max}=3.6\pm0.5 \mu$m/s and $K_M = (3.7\pm2) 10^2 \mu$M. (B) Measured activity  and shear viscosity of the AN as a function of ATP concentration. Error bars are the standard deviation of the mean for $v$, confidence bands for the fitted parameters $\alpha$ and $\eta$, and estimated confidence band from experimental observations for [ATP].
}
	\label{SFig:ATP}
\end{figure*}

For each pillar, we write an equation for each of the two spatial components in the force balance condition, $F_i = F^{def}_i$, which depends on $\alpha$ and $\eta$ (see SI  Sec. \ref{sec:balance}).  We find the optimum values for $\alpha$ and $\eta$ that more closely satisfy the force balance equations obtained for a set including dozens of pillars for different experimental realizations performed under the same experimental conditions (see SI Sec. \ref{sec:balance}). The results of our measurements are shown in Fig. \ref{fig:balance}B, where $\alpha$ and $\eta$ are plotted as a function of the average speed of the active flow. The latter is used here as a reliable proxy for the system's activity. A calibration curve of speed vs. ATP concentration is presented in Figure \ref{SFig:ATP}A, which allows to plot $\alpha$ and $\eta$ as a function of ATP concentration (Fig. \ref{SFig:ATP}B). Our results show that $\alpha$ is around $2\;10^{-7}$Pa m, only weakly dependent on activity, while $\eta$ is in the range $(4 - 10)\;10^{-6}$ Pa s m, revealing a shear thinning nature of the active material, as previously reported \cite{Rivas20}.

We have compared our direct measurement of $\eta$ with an early strategy reported by Mart\'{\i}nez-Prat et al. \cite{Martinez21} based on a model for the kinetic energy spectra, and have found excellent agreement (see SI Sec. \ref{sec:spectra} and Figure \ref{SFig:spectrum}).  Moreover, earlier theoretical studies predicted a scaling relationship between the activity parameter and the average speed of the form $v^2 \sim \alpha$ \cite{Giomi15}, which we have validated with our experimental data (see inset in Fig. \ref{fig:balance}D). We have also verified that our measurements agree with the theoretical expectation that $\alpha/\eta$ is related to the inverse of the intrinsic time scale of the AN, which we can extract from the average vorticity, $\omega$. Indeed, Fig. \ref{fig:balance}D suggests that these two parameters are proportional, $\omega\sim\alpha/\eta$ \cite{Giomi15,Joshi22}.

\begin{figure*}[h]
	\centering
	\includegraphics[width=0.9\textwidth]{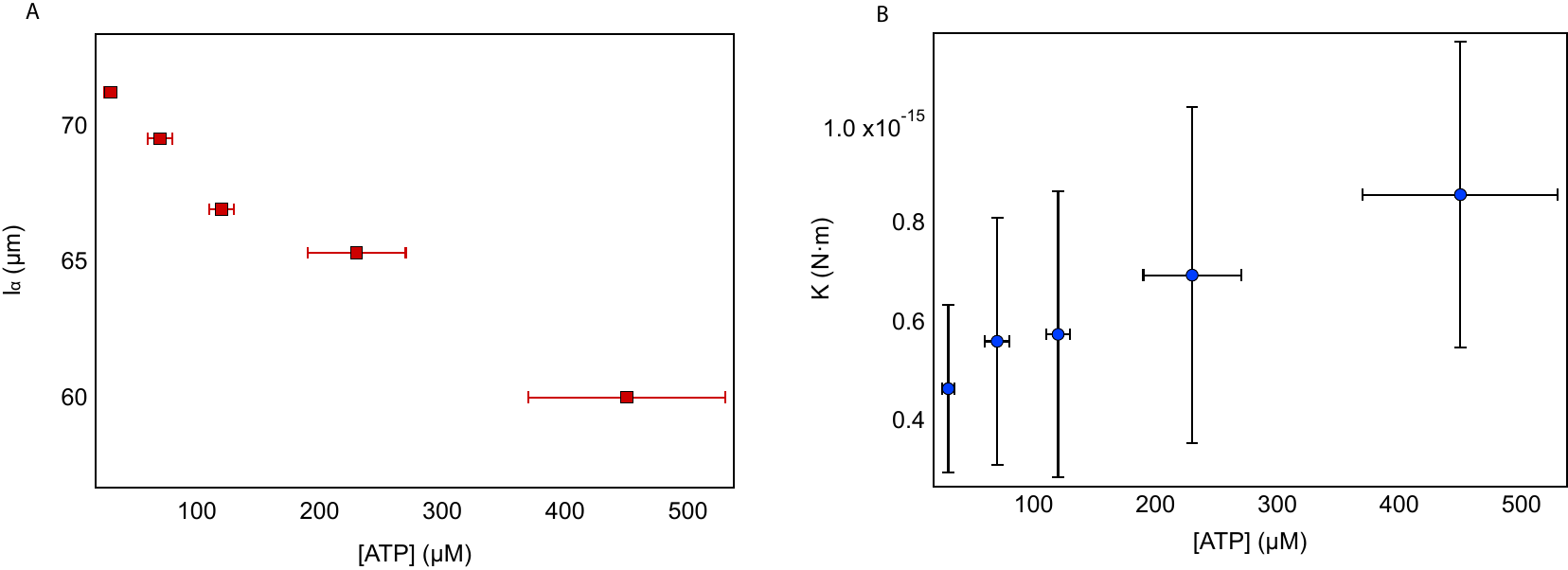}
	\caption{(A) Active length scale, $l_\alpha$, estimated from the defect density for different ATP concentrations. (B) Estimated values for the elastic constant of the active nematic, $K$, as a function of ATP concentration obtained from $l_\alpha$ and $\alpha$. Error bars are estimated confidence band from experimental observations for [ATP], and propagated errors from $l_\alpha$ and $\alpha$ for $K$.}
	\label{SFig:K}
\end{figure*}

Having obtained reliable values for $\alpha$ and $\eta$ at different activities, we can estimate a third material parameter, $K$, which encodes the bending rigidity of the active filaments. For this purpose, we resort to the scaling relation for the active length scale, $l_\alpha^2 \sim K/\alpha$ \cite{Doost18}. In our experiments, we estimate $l_\alpha \sim n^{-1/2}$, where $n$ is the average defect density, from which $K \sim \alpha/n$ can be estimated for each concentration of ATP (Fig. \ref{SFig:K}).  Our results suggest an average value $K \simeq 6\;10^{-16}$N m, with a marginal increase with the concentration of ATP, indicating that microtubule bundling may be enhanced by ATP.

\begin{figure}[t]
	\centering
	\includegraphics[width=\columnwidth]{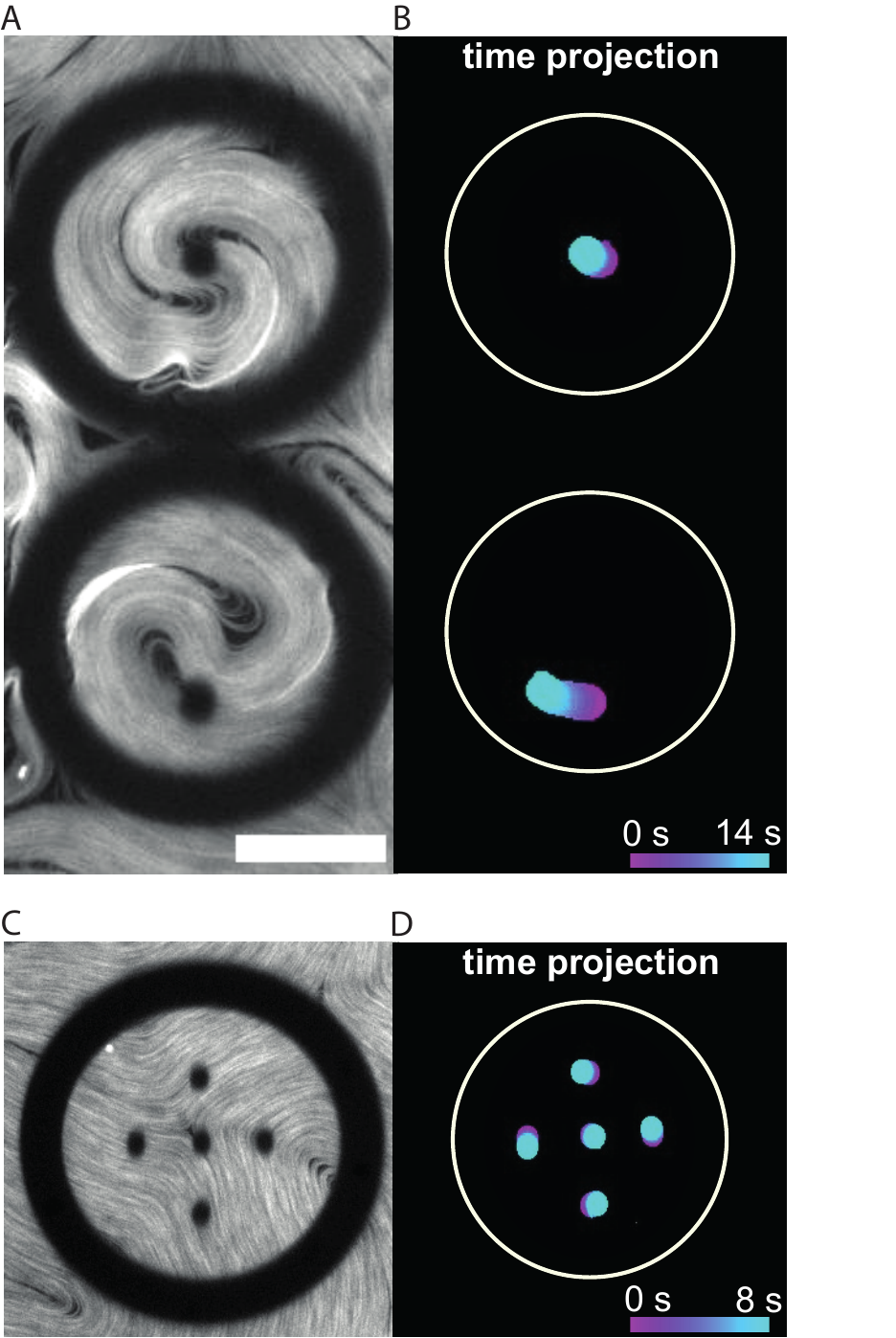}
	\caption{Mixed rigid and compliant objects for probing forces in confined AN arrangements. (A) Two wide rings impose rigid confinement on the AN layer, leading to the reorganization of the active flow into a pair of rotating $+1/2$ defects. Once the flow is steady, we imprint additional compliant columns, either at the center (top panel) or off-centered (botom panel) with respect to the rigid ring (see also SI Video \ref{SMovie:offcentered}). (B) Compliant columns deformation during the first 14 seconds after printing. (C) A rigid ring and a set of flexible columns are simultaneously imprinted around a $-1/2$ defect using 1.5 s of irradiation (see also SI Video \ref{SMovie:negative}). (D) Column deformation during the first 8 s after printing. The scale bar is 50 $\mu$m.
}	
\label{fig:confined}
\end{figure}

Having quantified the material parameters that determine the main local stress of the AN, the presented method enables us to quantitatively map, for any AN configuration, the resultant force density field, obtained as $f_i \simeq 2\eta \partial_j\varepsilon_{ij} -\alpha \partial_j Q_{ij}$. Note that here we have also neglected the contribution from the AN elasticity, as discussed above. In Fig. \ref{fig:forcemaps} we apply this result to map the total force density field around a $+1/2$ (Fig. \ref{fig:forcemaps}A) and a $-1/2$ (Fig. \ref{fig:forcemaps}B) defect. The force patterns for each configuration, respectively parabolic and triangular, are consistent with previously reported theoretical results \cite{Doost18,Serra23}. Experimental force patterns show that $+1/2$ defects self-propel from tail to head dragging the surrounding material. Conversely, we see that negative defects balance the local forces acting in their surroundings, leading to no net propulsion and exhibiting advected motion only.

\begin{figure*}[h]
	\centering
	\includegraphics[width=\textwidth]{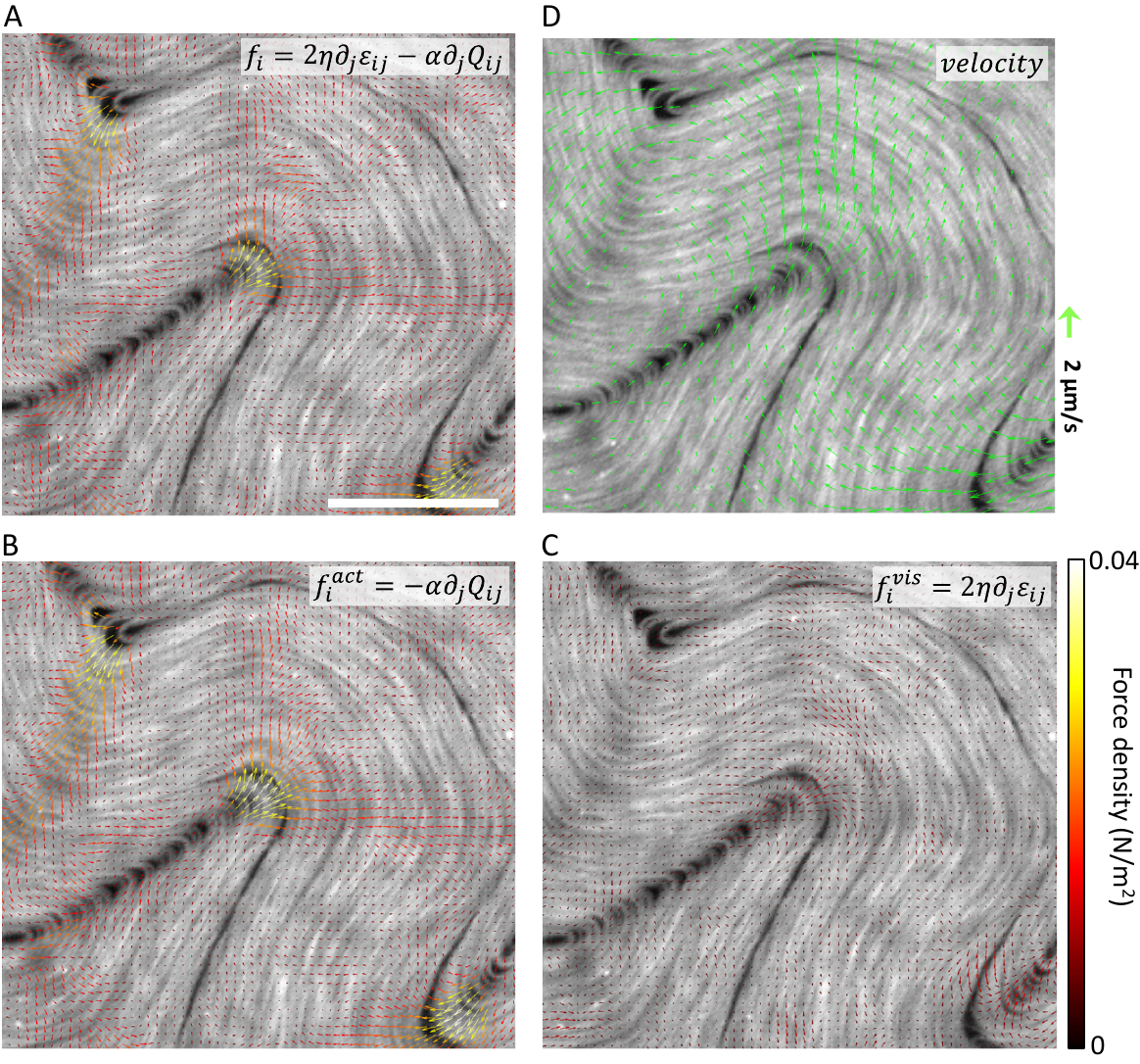}
	\caption{Maps of the force density surrounding a $+1/2$ defect in an AN layer with ATP = $50\pm10$ $\mu$M. Different panels represent the total AN force density (A), the active component of the force density (B), and the viscous component of the force density (C). For comparison, the map of the active flow velocity is included in panel D. The length of the arrows corresponds to the speed, as indicated by the speed scale. The scale bar is $50\;\mu$m.}
	\label{SFig:force_maps}
\end{figure*}

It is interesting to compare the relative strengths of the two dominant force contributions in the AN layer in the absence of obstacles. To this purpose, we have resolved the force density field into its active and viscous components near a $+1/2$ defect (SI Fig. \ref{SFig:force_maps}). The balance in this case is dominated by active forces, contrary to what was observed when computing the force balance for column deflection. In that case, viscous drag arising from no-slip boundary conditions on the hydrogel cylindrical surface resulted in strong forces in the direction of the average flow. Without obstacles, viscous force density is relatively weak and it opposes the self-propelling flows generated by the local active stress.

The propelling effect from  accumulated internal stress within aligned microtubule bundles is best illustrated during the unbinding of defect pairs of opposite topological charge, one of the most distinctive events in the dynamics of the quasi-two-dimensional ANs. One of this events is illustrated in Fig. \ref{fig:forcemaps}C, where the total force density increases at the tip of an unstable bending region. This promotes its growth, until a $+1/2$ defect unbinds, being self-propelled by a strong force density at its tip, while a $-1/2$ defect is left behind, surrounded by a stagnating force density field. The complementary scenario, namely that of defect annihilation, is shown in Fig. \ref{fig:forcemaps}D. The $+1/2$ defect is propelled by a strong force field acting on its tip towards a nearly stationary $-1/2$ defect. Recombination leads to a locally defect-free region where the force field vanishes.

The versatility of the employed method allows us to polymerize a combination of rigid and compliant inclusions.  This opens the door to measuring the force fields generated in confined AN arrangements. Some examples are presented in Fig. \ref{fig:confined}. Fig. \ref{fig:confined}A and B includes a system of two rotating positive defects created after active flows adapt to sudden ring confinement. Stresses are probed with one compliant column centered in one case, and off-center in the other. Column bending in each situation clearly reflects the stress pattern created by the AN, with small deformations near the center (top) and large deformations in the off-centered case (bottom). Fig. \ref{fig:confined}C and D explore, in a similar fashion, a $-1/2$ defect captured within a rigid ring. Here, the AN is prone to destabilization under confinement. An array of five compliant pillars seeded in a centered-square arrangement exhibit an inappreciable deformation of the central probe and small, albeit measurable, deformations of the outer columns that indicate a torque pattern.

The reported strategy of hybridizing the microtubule/kinesin system with a photopolymerizable monomer not only deepens significatively our understanding of the constitutive characteristics of this active gel but opens side by side endless possibilities to control the active nature of this particular and other similar materials.  The incorporation of active materials with externally-tunable characteristics into microchips represents a significant step forward in the design of future micro-machines built on the principles of active matter.\\

\textbf{Data availability}.
All data is available form the corresponding author upon reasonable request.

\acknowledgments
The authors are indebted to the Brandeis University MRSEC Biosynthesis facility for providing the tubulin. We thank M. Pons, A. LeRoux, and G. Iruela (Universitat de Barcelona) for their assistance in the expression of motor proteins.  I.V. acknowledges funding from Generalitat de Catalunya though a FI-2020 PhD. Fellowship. P.G. acknowledges
support from Generalitat de Catalunya through the Beatriu de Pin´os program (grant number 2020 BP 00248). I.V., J.I.-M., and F.S. acknowledge funding from MICINN/AEI/10.13039/501100011033 (Grant No. PID2019-108842GB-C22). Brandeis University MRSEC Biosynthesis facility is supported by NSF MRSEC 2011846. The authors acknowledge helpul discussions with C. S. and N. I-M.

\bibliography{refactive}

\clearpage

\setcounter{page}{1}
\setcounter{section}{0}

\beginsupplement
\nolinenumbers

\onecolumngrid
\begin{center}
\bf{SUPPLEMENTARY MATERIAL}
\end{center}

\vspace{1cm}

\begin{center}
{\bf Probing active nematics with in-situ microfabricated elastic inclusions}\\
\vspace{5mm}
V\'elez-Cer\'on \sl{et al.}
\end{center}

\twocolumngrid

\section{Methods}
\label{sec:Methods}

\subsection{Polymerization of microtubules}

Microtubules ($\sim 1\;\mu$m) were polymerized from heterodimeric ($\alpha$,$\beta$)-tubulin from bovine brain (Brandeis University Biological Materials Facility). Tubulin was incubated at 37 $^{\circ}$C for 30 min in aqueous M2B buffer (80 mM Pipes (piperazine-N,N'-bis(2-ethanesulfonic acid)), 1 mM EGTA (ethylene glycol-bis($\beta$-aminoethyl ether)-N,N,N',N'-tetraacetic acid), 2mM MgCl$_2$) (Sigma; P1851, E3889 and M4880, respectively) prepared with Milli-Q water and supplemented with the antioxidant agent dithiothrethiol (DTT) (Sigma 43815) and non-hydrolysable guanosine triphosphate (GTP) analog GMPCPP (guanosine-5'-[($\alpha$,$\beta$)-methyleno]triphosphate) (Jena Biosciences, NU-405) up to a concentration of 1 mM and 0.6 mM, respectively. To allow the observation under fluorescence microscopy, 3\% of the total tubulin concentration (8 mg ml$^{-1}$) was labelled with Alexa 647. Afterward, the mixture was annealed at room temperature for 4 h and kept at -80 $^\circ$C until use.

\subsection{Kinesin expression}

The kinesin used in the experiments was the heavy-chain kinesin-1 from Drosophila melanogaster truncated at residue 401, fused to biotin carboxyl carrier protein (BCCP), and labelled with six histidine tags (K401-BCCP-6His). This was expressed in Escherichia coli using the plasmid WC2 from the Gelles Laboratory (Brandeis University) and purified with a nickel column. After dialysis against 500 mM imidazole aqueous buffer, kinesin concentration was estimated by means of absorption spectroscopy. Finally, the kinesin was stored in a 40\% (wt/vol) aqueous sucrose solution at -80 $^\circ$C until use.

\subsection{Assembly of the microtubule-based active mixture}

Biotinylated kinesin motor proteins were incubated on ice for 30 min with tetrameric streptavidin (Invitrogen; 43-4301) at the specific ratio of 2:1 to obtain kinesin-streptavidin motor clusters. The standard preparation of the active mixture consisted of a M2B aqueous solution that contains ATP (adenosine 5'-triphosphate) (Sigma; A2383), the motor clusters, microtubules, and the depleting agent PEG (poly(ethylene glycol)) (20 kDa, Sigma;95172). To maintain a constant concentration of ATP during the experiments, an enzymatic ATP-regeneration system was added, consisting on phosphoenolpyruvate (PEP) (Sigma; P7127) that fueled the enzyme pyruvate kinase/lactate dehydrogenase (PKLDH) (Sigma; P0294) to convert ADP (adenosine 5'-diphosphate) back to ATP. The mixture also contained several antioxidants to avoid photobleaching during the fluorescence imaging: DTT, Glucose Oxidase (Sigma; GT141), catalase (Sigma; C40), Trolox (6-hydroxy-2,5,7,8-tetramethylchroman-2-carboxylic acid) (Sigma; 238813), and glucose (Sigma; G7021). The constituents of the photopolymerizing hydrogel were incorporated in solid form to the active mixture. The former consisting on the photoinitiator lithium phenyl-2,4,6-trimethylbenzoylphosphinate (LAP, TCI; L0290) and the monomer 4-ArmPEG-Acrylate 5 kDa (4PEG5k, Biochempeg; A44009-5k). The final concentrations of each reagent used in the preparation of the active material is shown in Table \ref{table:Conc}.

\begin{table}[t]
    \begin{tabular}{|l|l|c|l|}
    \hline
        \textbf{Compound} & \textbf{Buffer} & \textbf{Final Conc.} & \textbf{Units}  \\ \hline
        PEG (20 kDa) & M2B & 1.54 & \% w/v  \\ \hline
        PEP & M2B & 25.68 & mM  \\ \hline
        MgCl$_2$ & M2B & 3.12 & mM  \\ \hline
        ATP & M2B & 60 - 296 & $\mu$M  \\ \hline
        DTT & M2B & 5.21 & mM  \\ \hline
        Streptavidin & M2B & 0.01 & mg/mL  \\ \hline
        Trolox & Phosphate & 1.93 & mM  \\ \hline
        Catalase & Phosphate & 0.04 & mg/mL  \\ \hline
        Glucose & Phosphate & 3.20 & mg/mL  \\ \hline
        Glucose Oxidase & Phosphate & 0.21 & mg/mL  \\ \hline
        PK & Original & 25.01 & u/mL  \\ \hline
        LDH & Original & 24.91 & u/mL  \\ \hline
        Kinesin & Original & 0.08 & mg/mL  \\ \hline
        Microtubules & Original & 1.85 & mg/mL  \\ \hline
        LAP & Mixture & 0.226 & \% w/v  \\ \hline
        4PEG5k & Mixture & 4.52 & \% w/v  \\ \hline
    \end{tabular}
    \caption{Composition of all solutions (including buffer used for their preparation), and concentration of the different species in the final mixture. Acronyms used in this table are: PEG (Poly-ethylene glycol); PEP (Phosphoenol pyruvate); ATP (Adenosin triphosphate); PK (Pyruvate Kinase); LDH (Lactic Dehydrogenase); DTT (1,4-dithiothreitol); 4PEG5k (4-ArmPEG-Acrylate 5 kDa); LAP (lithium phenyl-2,4,6-trimethylbenzoylphosphinate). The two latter compounds where directly incorporated to the active mixture. M2B buffer: 80 mM PIPES (piperazine-N,N'-bis(2-ethanesulfonic acid)) pH 6.8, 2 mM MgCl$_2$ 1 mM EGTA (egtazic acid). Phosphate buffer: 20 mM Phosphate buffer (6.68 mM KH$_2$PO$_4$, 12.32 mM K$_2$HPO$_4$) pH 7.2;  Original: species is obtained already dissolved in its custom buffer. }
    \label{table:Conc}
\end{table}

\subsection{Active nematic cell}

The experiments were carried out using flow cells with a channel of 3-5 mm width and 50 $\mu$m height. The cell was composed of a bioinert and superhydrophilic polyacrylamide-coated glass and a hydrophobic Aquapel-coated glass, separated using a 50 $\mu$m thick double-sided tape. The cell was first filled by capillarity with fluorinated oil (HFE7500; Fluorochem 051243) which contains 2\% of a fluorosurfactant copolymer (RanBiotechnologies, 008 Fluorosurfactant). The active material was subsequently introduced in the cell by capillarity. To avoid evaporation, the cell was sealed using petroleum jelly. This preparation resulted in the approximate thickness for each phase (as determined using fluorescence confocal microscopy): 10 $\mu$m for the oil layer, 5-10 $\mu$m for the AN layer, and 30-35 $\mu$m for the aqueous subphase.

\subsection{Observation of the active nematic}

The observation of the active nematic layer was carried out by means of fluorescence microscopy. We used a Nikon Eclipse Ti2-e equipped with a white LED source (Thorlabs MWWHLP1) and a Cy5 filter cube (Edmund Optics). Images were captured using an ExiBlue CCD camera (QImaging) operated with the open-source software ImageJ $\mu$-Manager.

\subsection{Hydrogel Polymerization}

\begin{figure}[t]
	\centering
	\includegraphics[width=0.7\columnwidth]{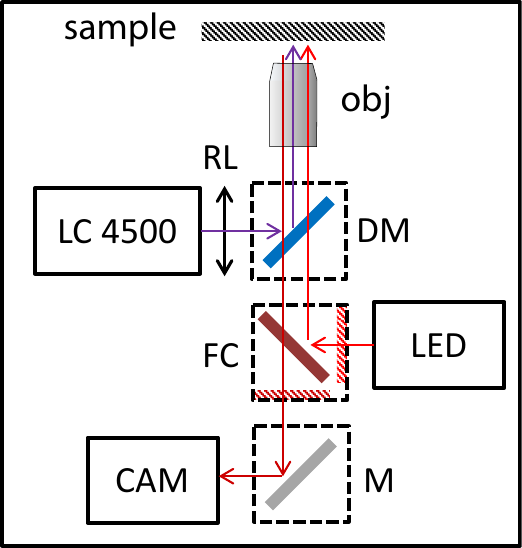}
	\caption{Optical setup as described in the text. }
	\label{SFig:optics}
\end{figure}

The fluorescence microscope was modified in order to incorporate a digital micromirror device (DMD) UV light projector (Texas Instruments LightCrafter 4500, with a 2W 385nm LED; EKB Technologies, Ltd), see Fig. \ref{SFig:optics}. Projected patterns are incorporated into the light path of the inverted microscope by means of a collimating lens (f = +150 mm) and a 505 nm dichroic mirror (Thorlabs DMLP505R), and are focused on the sample by means of the microscope objectives, reaching a lateral resolution up to a few microns. The DMD projector is connected as an external monitor to a computer, thus enabling real time control of the projected patterns using MS-PowerPoint slides. For the experiments shown in this work, the hydrogel was polymerized using an x20 objective, resulting in a light power density of 3.1 W cm$^{-2}$. The resulting columns have a height of $40\;\mu$m, of which 5-10 $\mu$m are embedded in the AN layer.

The rigidity of the hydrogel structures can be adjusted by a combination of the irradiation time and the power of the led light source that feeds the DMD device. Since the AN features self-sustained flows that drag the water subphase, polymerization must be performed quickly to prevent the formation of distorted shapes, as the incipient hydrogel is advected by the active flows. In practice, we require that the maximum flow advection during polymerization is under 2 $\mu$m s$^{-1}$, which limits ATP concentration to be below 450 $\mu$M. With the used concentration of monomer and initiator, we find that the polymerization time must be in the range 750 - 1250 ms for soft and 1500 - 3000 ms for rigid structures. As an alternative to adjusting the polymerization time, the lateral compliance of the structures can be set by controlling their in-plane cross section.

\subsection{Sample characterization}

The director field of the active nematic layer was obtained using a coherence-enhanced diffusion filtering (CEDF) MatLab code \cite{Ellis20} applied on fluorescence images. Tracer-free velocimetry analysis of the active nematic was performed with the public MatLab App PIVlab. Further analysis of director and velocity field data was performed with custom-written MatLab scripts.

\subsection{Young’s modulus determination}

The hydrogel Young’s modulus was determined in-situ, in a flow cell prepared as explained above, filled by capillarity with the buffer solution and gellifying agents but without motor proteins and microtubules. Cylindrical columns were polymerized using the protocol described above and were deformed by flowing liquids of known viscosity by capillarity. We used methylcellulose (Sigma; M0512) solutions of concentrations between 0.18 and 0.54\% (w/vol) prepared in M2B buffer. Their viscosity was measured using a Brookfield Ametek DVE Viscosimeter with a Yula-15 spindle. To measure the average flow speed, a 5\% (vol/vol) concentration of fluorescent particles of diameter 1 $\mu$m (Bangs Laboratories; FCDG006) were incorporated into the solution, and average speeds where determined by tracking individual particles using ImageJ.

\section{Data Analysis}

\subsection{Force balance}
\label{sec:balance}
The viscosity, $\eta$, and the activity parameter, $\alpha$, of the active nematic
were determined by analysing the in-plane ($xy$) resolved deflection, $\delta_i$, of the hydrogel cantilevers. Invoking basic concepts in mechanics of materials \cite{Hibbeler}, the differential equation for the deflection
can be formulated in terms of the bending moment, $M$, of a cylindrical pillar, its second moment of inertia, $I=\pi r^{4}/4$, where $r$ denotes the radius of the beam, and the Young's modulus of the material, $E$,

\begin{equation}
EI\frac{d^{2}\delta_i}{dz^{2}} =  M_i(z),
\label{deflection}
\end{equation}
where the index $i$ refers to each of the two spatial dimensions. The cantilevers, of length $L$, were fixed on the substrate. Their bending moment is equal to the deflection force multiplied by the distance from the position where it is applied, $M_i(z) = {F_{i}}^{def}z$.

By integrating Eq. \ref{deflection} with the corresponding boundary conditions $\delta_i (z=L)= \frac{d\delta_i}{dz}|_{z=L}=0$, the final relationship between the deflection and the applied force components reads,

\begin{equation}
F_{i}^{def}=\frac{3\delta_i EI}{L^{3}} = \frac{3\pi}{4}\frac{\delta_i Er^{4}}{L^{3}}
\end{equation}
expressed in terms of the geometrical parameters of a beam and its Young's modulus .


This deflection force is balanced with the resultant from adding the force directly exerted by the active fluid, and those indirectly arising in the driven contact fluids, which are hydrodynamically coupled to the AN layer.
The force acting on the submerged columns can be expressed in terms of the total stress acting on them, $\sigma_{ij}$, as \begin{equation}
F_i = \iiint_{V_c} \sum_{j=1,2} \partial_j \sigma_{ij} dV = \iint_{S_c} \sum_{j=1,2} \sigma_{ij}n_jdS.
\end{equation}
Here, $V_c$ spans the volume of the cylinder, $S_c$ spans the exposed area of the cylinder (lateral and free circular surface), and $\hat{n}$ is the unit vector normal to the local surface. We split this computation into three parts,
\begin{equation}
\begin{split}
F_i & = F_i^{oil} + F_i^{AN} + F_i^{water} \\
& = \iint_{S_o} \sigma^o_{ij}n_jdS + \iint_{S_{AN}} \sigma^{AN}_{ij}n_jdS + \iint_{S_w} \sigma^w_{ij}n_jdS,
\end{split}
\end{equation}
where the indices $o$, $AN$, and $w$ stand, respectively, for the oil layer in contact with the free top cylinder surface, the cylindrical region of height $\Delta z$ where the cylinder is submerged in the AN layer, and the layer of aqueous subphase that supports the active layer. We justify below that, given the large viscosity contrast between the AN layer and the aqueous phase and the used oil, the direct contribution of the AN is clearly dominant within the force balance.

The force exerted by the AN layer on the cylindrical region of thickness $\Delta z$ is computer as
\begin{equation}
\begin{split}
F^{AN}_i & = \iint_{S_{AN}} \sum_{j=1,2} \sigma^{AN}_{ij}n_jdS = \oint \sum_{j=1,2} \sigma^{AN}_{ij}n_j dl\;\Delta z \\
& = \oint \sum_{j=1,2} \sigma^{AN,2d}_{ij}n_j dl,
\end{split}
\end{equation}
where the rightmost integration is performed along the circular edge of the cylinder free surface. We implicitly assume that $\sigma^{AN}_{ij}$ is invariant across the thickness of the AN layer. As a result, we have defined the two-dimensional AN stress tensor, $\sigma^{AN,2d}_{ij} = \sigma^{AN}_{ij}\Delta z$, with which we implement the force balance without the need for a precise knowledge of the thickness of the AN layer.

The 2D stress tensor is further decomposed in terms of the dominant viscous and active stresses,
$\sigma^{AN,2d}_{ij} = \sigma_{ij}^{vis} + \sigma_{ij}^{act}$,
respectively defined as $\sigma_{ij}^{vis} = 2\eta \varepsilon_{ij}$, and
$\sigma_{ij}^{act} = - \alpha Q_{ij}$. In these last expressions, $\varepsilon_{ij}$ denotes the  strain rate
tensor, and $Q_{ij}$ is the traceless nematic tensor order parameter, both in their two-dimensional forms.

Under the hypothesis of no-slip boundary conditions on the hydrogel column surface, the inclusion of the latter inside the active material produces a drastic change in the velocity field next to them. To accurately compute the strain rate tensor it is thus necessary to achieve a substantial spatial resolution of the velocity field around the obstacles. Since this condition is not totally guaranteed from our PIV analysis, we resorted to estimating the viscous drag using Lamb's equation \cite{Landau},

\begin{equation}
F_{i}^{drag} = 4\pi\epsilon\eta^{3D}v_i\Delta z = 4\pi\epsilon\eta v_i,
\label{eq:drag}
\end{equation}
in terms of the average velocity just before printing the pillars, and the viscous drag coefficient, $\epsilon$. For the very low Reynolds number in our experiments the latter is estimated as \cite{Landau}

\begin{equation}
\epsilon = {\lbrack 0.5 - \gamma - \ln\left( \frac{Re}{8} \right)\rbrack}^{- 1},
\label{eq:Lamb}
\end{equation}
where $Re$ is the Reynolds number, and $\gamma$ is the Euler constant. In our experiments we took $\epsilon=0.052$, corresponding to an averaged value of the Reynolds number $Re=3.2 \times10^{-8}$. The value of $L$ is considered to be 40 $\mu$m (see above), while $r$ varied between 5.2 and 9 $\mu$m. For each tested condition, experiments were repeated with different cylinder radii to achieve an average deflection between $r$ and $2r$, approximately, which resulted in the most reliable results for $\alpha$ and $\eta$.

For each column, one can thus write the force balance separating explicitly the distinctive linear contributions in $\eta$ and $\alpha$ as,

\begin{equation}
F_{i}^{def} = F_{i}^{drag} + F_{i}^{act} = F_{i}^{\eta}\eta + F_{i}^{\alpha}\alpha,
\label{eq:system}
\end{equation}
with $F_{i}^{\eta}=4\pi\varepsilon v_i$ and $F_{i}^{\alpha} = -\oint \sum_{j=1,2} Q_{ij}n_j dl$.
This equation was applied for each column, using data for the director and velocity  fields measured at the column location right before polymerization. In addition, we assumed that the cylindrical cantilevers reach mechanical equilibrium when they change noticeably their initial moving direction or when they reach a steady deflection value.

For each column we obtained two equations, corresponding to the two components of $\vec{F}^{def}$, as shown in Eq. \ref{eq:system}. Therefore, if we analyze $N$ columns, we will obtain $2N$ equations for the two unknowns $\eta$ and $\alpha$. This overdetermined system of equations was solved numerically finding the two parameters that minimize the sum of the squares of all the errors. In the different sets of performed experiments the number of analyzed cylindrical cantilevers varied between 37 and 118.

In this study, hydrogel columns where polymerized in a hexagonal lattice over the field of view, with no a priori consideration on the local configuration of the surrounding AN. This made some of the columns unsuitable for the analysis described above. Columns that were discarded satisfied one or more of the following conditions:

\begin{itemize}
\item
  Incorrect tracking. The contour of columns that were polymerized on top of AN defects could not be tracked with enough precision, as those regions are deprived of fluorescent microtubules. This makes them indistinguishable from hydrogel columns.
\item
  Regions with aligned filaments. In highly aligned AN regions, the Q tensor is practically uniform in space, and velocity tracking using PIV software cannot be performed with enough precision.
\item
  Highly distorted regions. In the area of influence of more than one AN defect, the Q tensor and the velocity change considerably over small distances. This made force quantification in those regions too sensitive to the precise location of cylinder boundaries, resulting in imprecise computation of the viscous and active force components.
\end{itemize}

\subsection{Drag force from the oil and water phases}
\label{sec:drag}
Self-sustained motion of the AN layer leads to flows on both the contacting oil layer and the underlying aqueous phase. Both fluid phases exerted drag forces on the polymerized cylindrical columns. In this section, we quantify the resulting forces and demonstrate that they can be discarded when compared to the forces exerted by the AN layer.

The oil phase can only produce drag on the free surface of the column. This drag force can be calculated using the classical result for drag on a disk \cite{Landau} that, assuming no-slip boundary conditions at the AN-oil interface, reads

\begin{equation}\label{oildrag}
 F_{oil}^{drag} = \frac{1}{2}(\frac{32}{3}\eta_{oil}rv).
\end{equation}

We compare this expression with Eq. \ref{eq:drag} to estimate the relative relevance of the drag by the oil layer,
\begin{equation}
   \frac{F^{drag}}{F_{oil}^{drag}} = \frac{3\pi\epsilon}{4}\frac{\eta}{\eta_{oil}r}.
   \label{eq:comparison}
\end{equation}
The oil used in the experiments has a viscosity of 1.24$\times10^{-3}$\,Pa s, and cylinders are a few microns in diameter. With $\eta$ of the order of $10^{-6}$\,Pa s m (see Fig. \ref{fig:balance}D main text), and $\epsilon$ a numerical factor of order $10^{-2}$, the numerator in Eq. \ref{eq:comparison} is around two orders of magnitude larger than the denominator, which indicates that drag due to the oil layer can be safely neglected.

On the other hand, for the drag force exerted by the water phase all along the column, we take the simplifying assumption that there is a linear velocity profile that falls to zero on the substrate where columns are attached,
\begin{equation}
   v_{water}(z) = v\frac{z}{L}.
\end{equation}
We, then, evaluate the deflection under the non-uniform load that results from applying Lamb's equation for the viscous drag to the above velocity profile. Integrating with the same boundary conditions indicated earlier, we obtain
\begin{equation}
 \delta_{water} = \frac{8}{15}\frac{\epsilon\eta_{water}vL^{4}}{Er^{4}},
\end{equation}
and, by comparison with the deflection due to the AN layer,
\begin{equation}
  \frac{\delta}{\delta_{w}} = \frac{\frac{4}{3\pi}\frac{F^{drag}L^{3}}{Er^{4}}}{\frac{8}{15}\frac{\epsilon\eta_{water}vL^{4}}{Er^{4}}} = \frac{5}{2\pi}\frac{4\pi\epsilon\eta v}{\epsilon\eta_{water}vL} = 10\frac{\eta}{\eta_{water}L},
\end{equation}
we find the contribution of the aqueous phase to the deflection of the hydrogel cylinders to be less than $1\%$ of the contribution by the AN layer. Hence, the effect of the aqueous subphase can be neglected.

\subsection{In-situ determination of the hydrogel Young's modulus}
\label{sec:Young}

\begin{figure*}[th]
	\centering
	\includegraphics[width=0.8\textwidth]{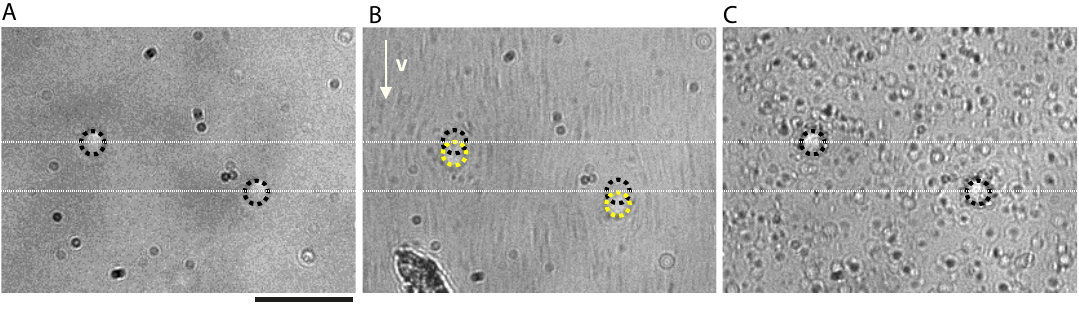}
	\caption{Calibration of the hydrogel rigidity. (A) Columns with $r = 5.4\;\mu$m are imprinted in a standard cell filled with M2B buffer and the precursors of the hydrogel. The circular contour of the columns is highligtd with black dashed lines in the brightfield images (B) The unpolymerized buffer is replaced with M2B buffer containing methylcellulose and fluorescent microparticles. Columns bend due to viscous drag. Their contour is highlighted with yellow dashed lines. (C) When flow stops, columns relax to their initial shape. Seeded microparticles can be observed in this panel. White dotted lines indicate the center of the circular cross section of the original columns at different times. Scale bar is 50 $\mu$m.}
	\label{SFig:calibration}
\end{figure*}

Calibration experiments employing the same protocol for hydrogel photopolymerization were performed for an \textit{in situ} determination of its Young's modulus (Fig. \ref{SFig:calibration}). For this purpose, we measured the deflection of cylindrical columns due to the viscous drag of an aqueous fluid of know viscosity that flows in the cell at a known velocity.

In this case, deflection is purely due to the viscous drag from the resultant contributions of the aqueous and oil phases present in the standard flow cell. The aqueous phase exerts its drag all along the body of the column. For simplicity, we will assume that a uniform average force per unit length $q$ is applied along the height of each column. We resort once again to Lamb's equation to obtain $q=4 \pi \epsilon \eta_{water} v$, where $v$ will be the average flow speed in the aqueous phase. On the other hand, the contribution from the oil phase is simply expressed in terms of the corresponding drag on the free column surface, as given in Eq. \ref{oildrag}. Thus, the final expression for the total bending moment reads,

\begin{equation}
 M(z) = q \frac{z^2}{2}+F_{oil}^{drag}z.
\end{equation}
Integrating the differential equation for the deflection (Eq. \ref{deflection}), we arrive at an explicit equation for the deflection written as,

\begin{equation}
\delta= \frac{3qL^4+8F_{oil}^{drag}L^3}{6\pi Er^4}.
\end{equation}

Different experiments were carried out with aqueous methylcellulose solutions, spanning a range of viscosities between 3.6 and 18.7 mPa s, flowing at velocities between 5 and 27$\,\mu$m s$^{-1}$. For each experiment we evaluated the corresponding Reynolds-number-dependent drag coefficient $\epsilon$. The final estimation for the Young's modulus evaluated this way is $E= 18.1\pm 1.5$ Pa.

\subsection{Active nematic viscosity estimated from kinetic energy spectra}
\label{sec:spectra}

In earlier work, Mart\'{\i}nez-Prat et al. (Nature Physics \textbf{15}, 362(2019)) estimated the AN viscosity by fitting an ad hoc model for the kinetic energy spectrum. In that work, boundary conditions for the oil layer in contact with the AN where different from the ones we have here. While Mart\'{\i}nez-Prat et al. used an open cell and, thus, no-slip boundary conditions were assumed at the oil/air interface, here our cells are closed, so we need to invoke no-slip boundary conditions when both the water subphase and the oil layer contact with the confining glass plates. This leads to a modified equation for the kinetic energy spectrum,

\begin{equation}
E(q) = \frac{Bq{R_{*}}^{4}e^{- q^{2}{R_{*}}^{2}/2}\lbrack I_{o}\left( \frac{q^{2}{R_{*}}^{2}}{2} \right) - I_{1}\left( \frac{q^{2}{R_{*}}^{2}}{2} \right)\rbrack}{{\lbrack q + \frac{\eta_{oil}}{\eta}{\coth}\left( qH_{oil} \right) + \frac{\eta_{w}}{\eta}\coth{\left( qH_{w} \right)\rbrack}}^{2}}.
\label{eq:spectrum}
\end{equation}

In this equation $q$ is the wave number, $\eta_{oil}$, $\eta_w$, and $\eta$ are the viscosities of the oil, water, and the active nematic layer; $H_{oil}$ and
$H_{w}$ are the thickness of the oil and water phases, respectively; $R^*$ is the mean vortex radius of the AN flow, $I_0$ and $I_1$ are modified Bessel functions;
$B = N\omega_{v}^{2}/(32\pi^{3}\mathcal{A)}$ is a prefactor related to the total enstrophy, where N is the average number of vortices over a
total system area $\mathcal{A}$, and $\omega_v$ is the average vorticity.

\begin{figure}[h]
	\centering
	\includegraphics[width=0.45\columnwidth]{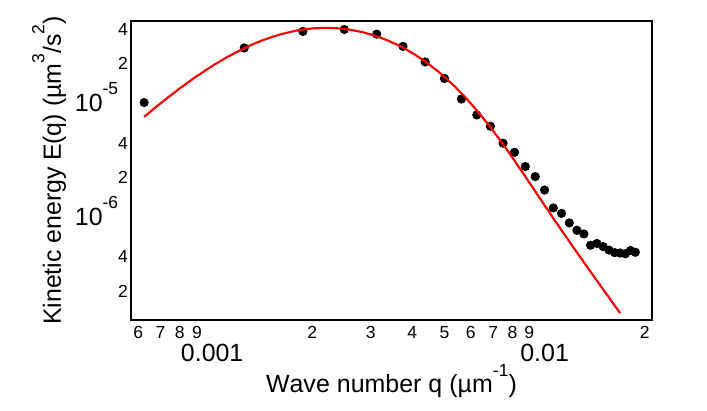}
	\caption{Kinetic energy spectrum of the AN layer with [ATP] = 450 $\mu$M. The solid line is a fit to a variant of the model proposed by Mart\'{\i}nez-Prat et al. that yield an estimation for the shear viscosity of the active material $1.3\;10^{-6}$ Pa s m.}
	\label{SFig:spectrum}
\end{figure}

As a check for the values of $\eta$ obtained in this work using direct rheological measurements, we have computed the kinetic energy spectrum for one experiment with an average speed around $1\;\mu$m/s and we have fitted the data to Eq. \ref{eq:spectrum} (Fig. \ref{SFig:spectrum}), yielding an estimation for the AN viscosity $\eta = (1.3 \pm 0.1) 10^{-6}$ Pa s m. This value is of the same order as the one we obtain in this work using in-situ microrheology.

\cleardoublepage

\onecolumngrid

\begin{center}
\bf{SUPPLEMENTARY FIGURES}
\end{center}

\cleardoublepage

\setcounter{figure}{0}

\renewcommand{\figurename}{Movie }

\onecolumngrid
\begin{center}
\bf{SUPPLEMENTARY MOVIES}
\end{center}

\begin{figure}[h]
	\centering
	\includegraphics[width=0.5\columnwidth]{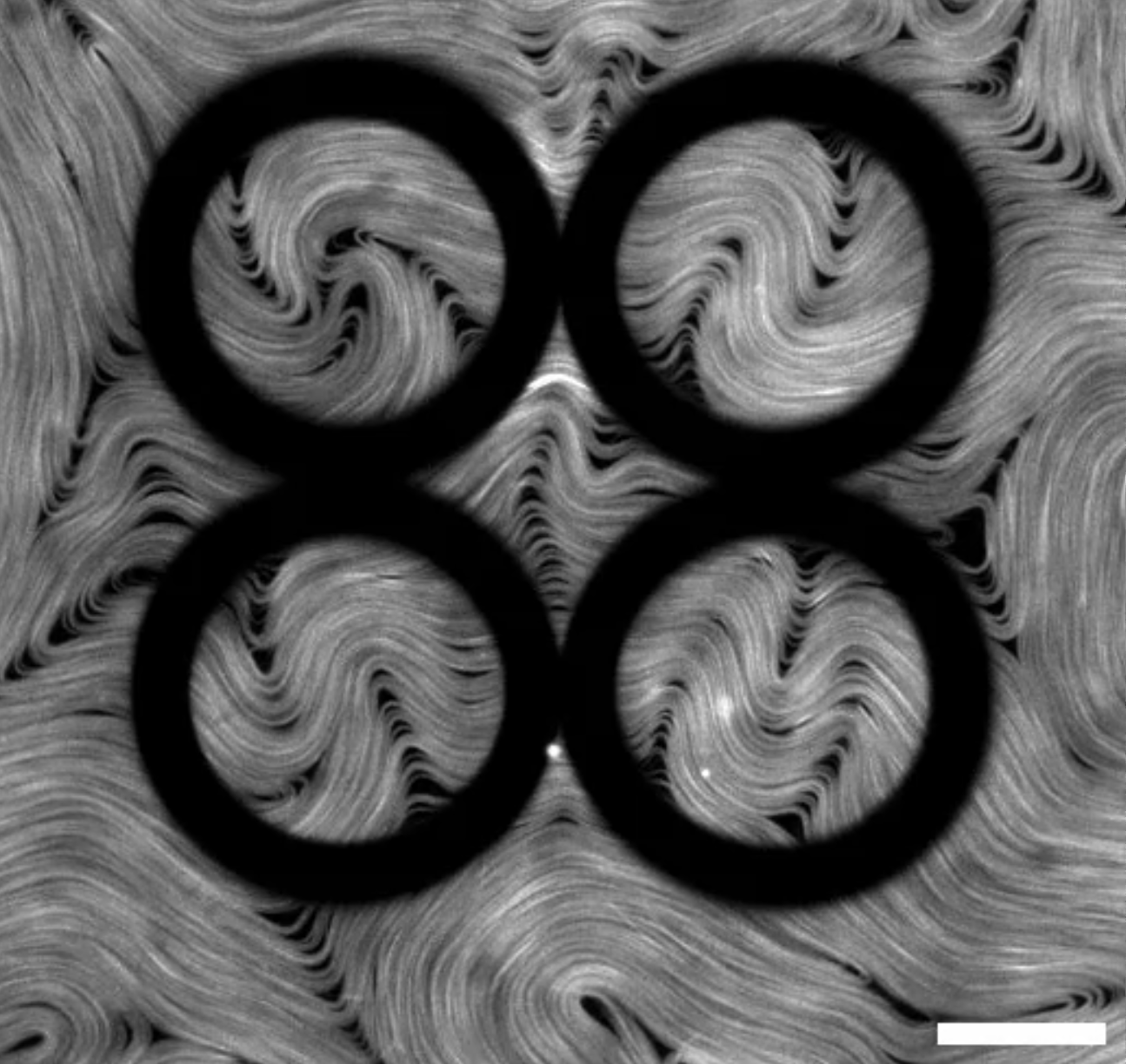}
	\caption{A set of four rigid rings that are in-situ polymerized in the AN cell. Scalebar, $50\;\mu$m.
}
	\label{SMovie:rings}
\end{figure}

\begin{figure}[h]
	\centering
	\includegraphics[width=0.5\columnwidth]{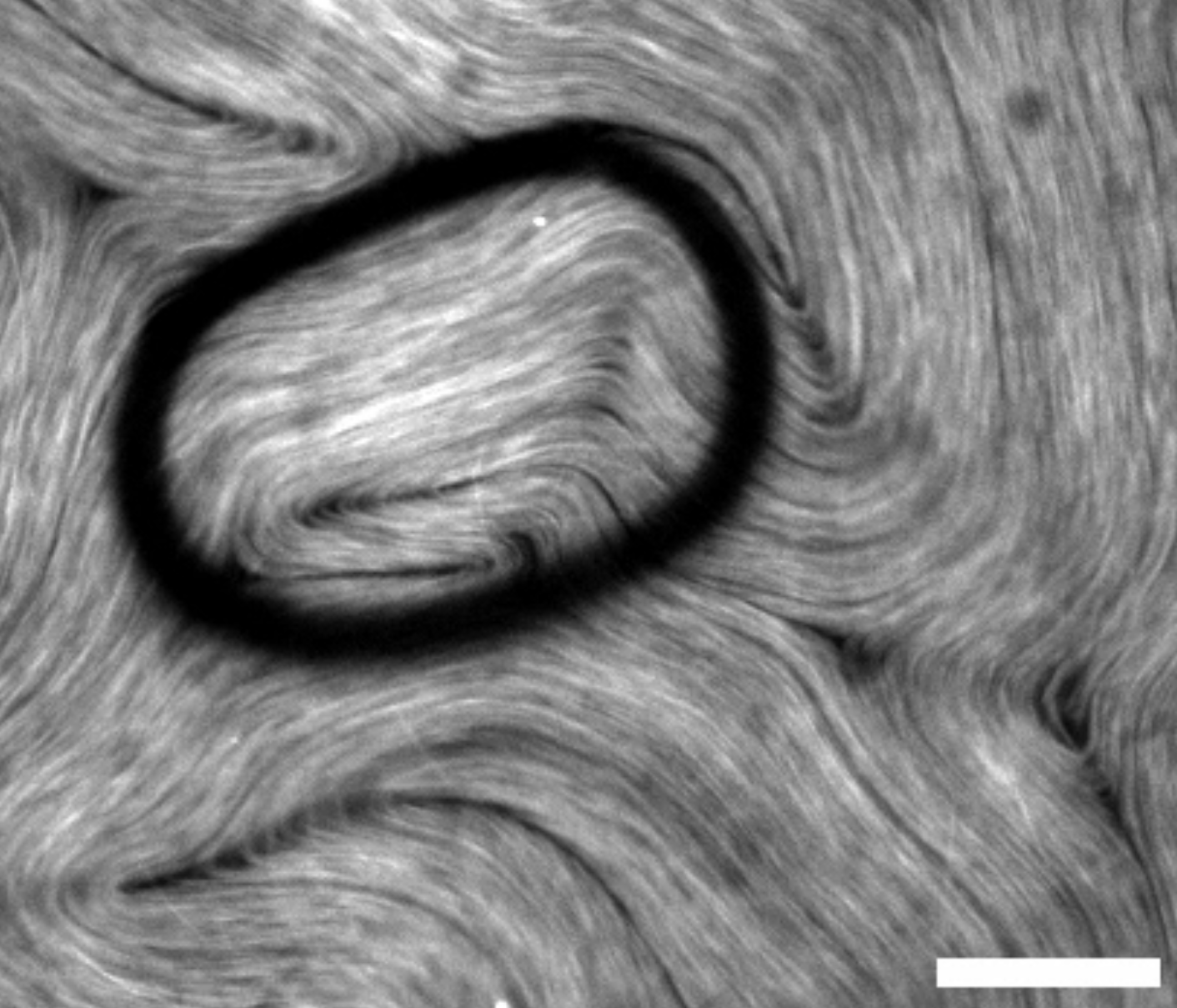}
	\caption{A flexible ring that has been in-situ polymerized in the AN cell. Scale bar, $50\;\mu$m.
}
	\label{SMovie:flexible}
\end{figure}

\begin{figure}[h]
	\centering
	\includegraphics[width=0.5\columnwidth]{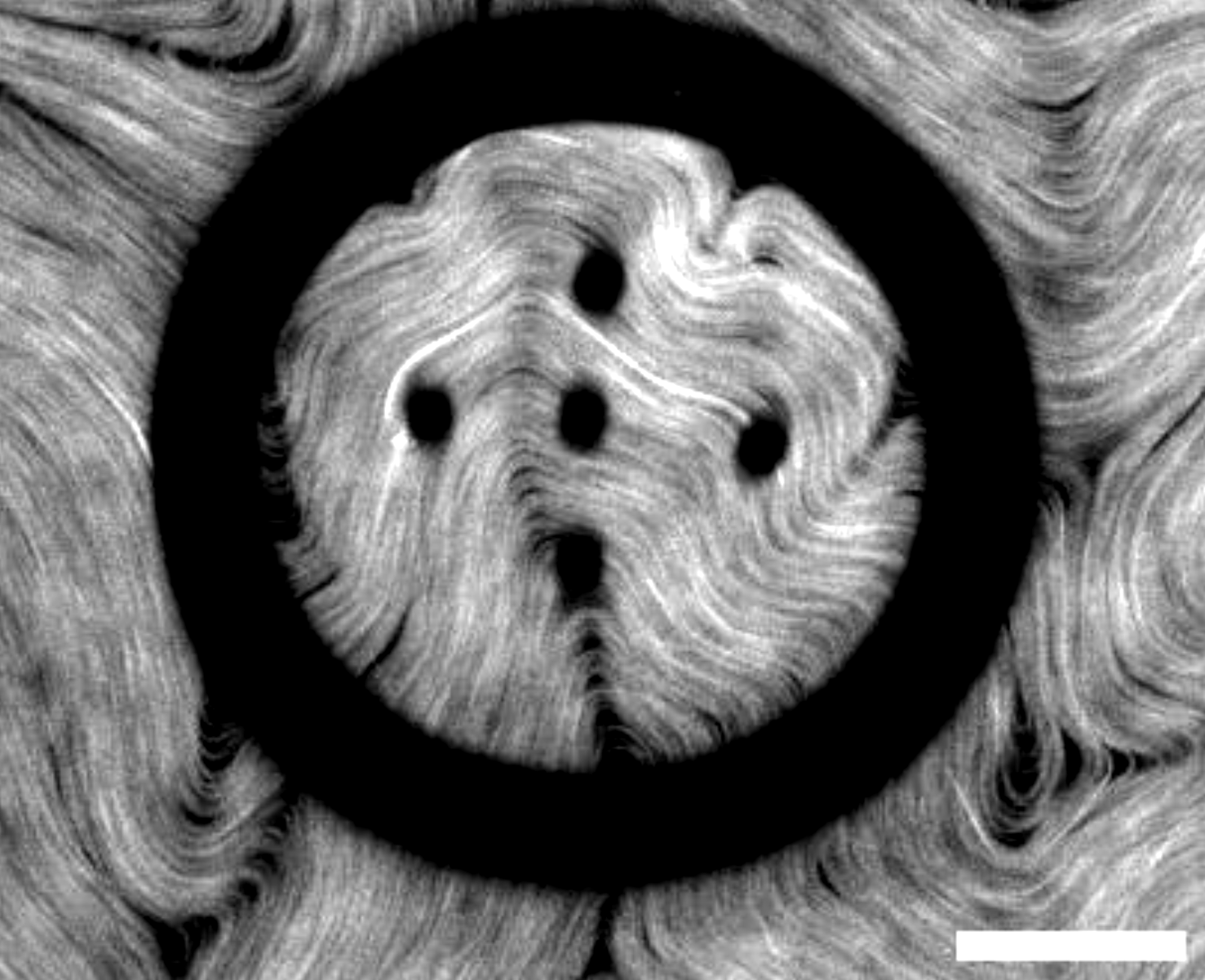}
	\caption{A hybrid combination of a rigid ring boundary with embedded soft columnar pillars, all built around a $+1/2$ defect. Scale bar, $50\;\mu$m.
}
	\label{SMovie:hybrid}
\end{figure}

\begin{figure}[h]
	\centering
	\includegraphics[width=0.5\columnwidth]{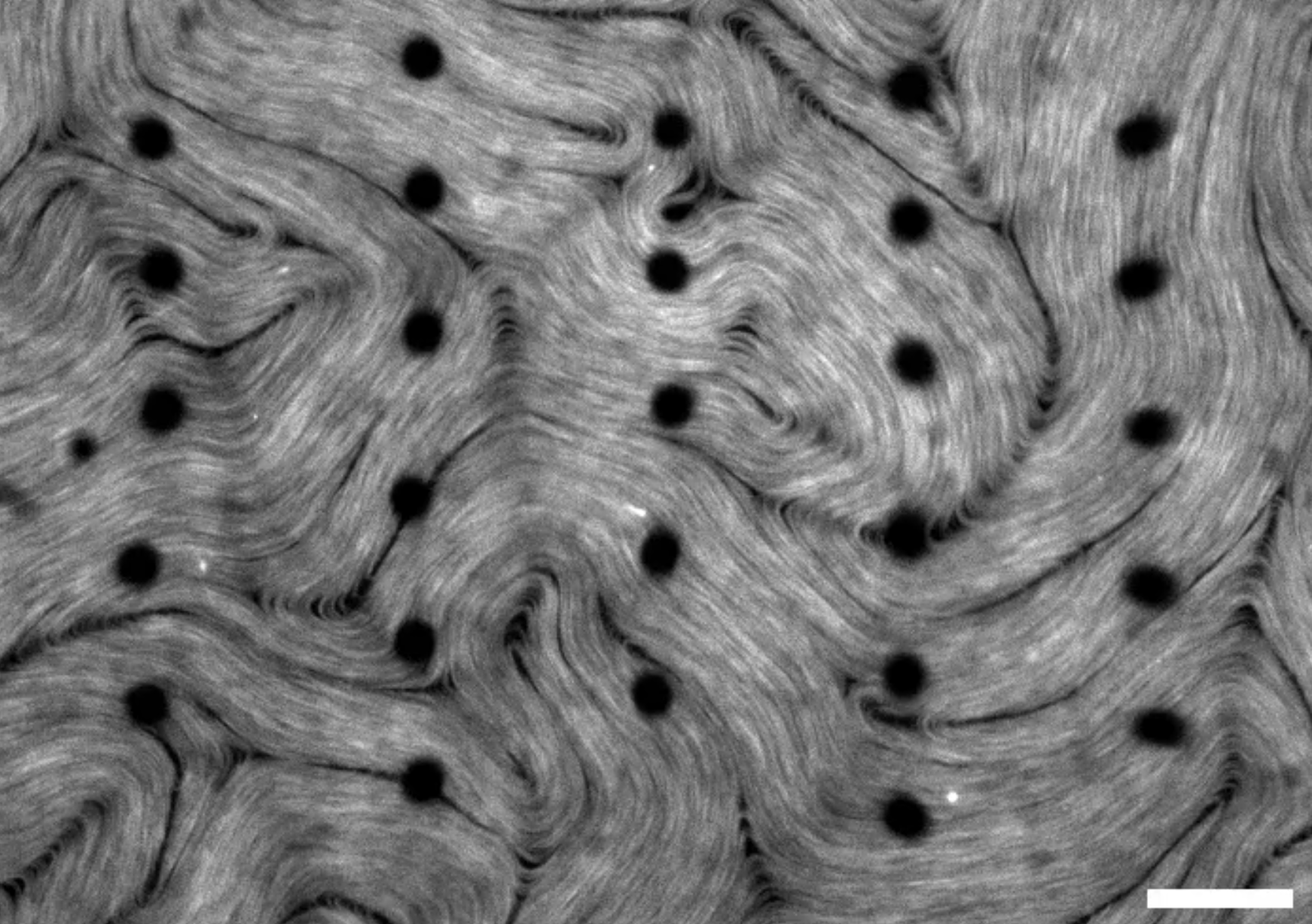}
	\caption{An array of soft pillars distributed in a hexagonal lattice to probe the microrheology of the AN layer. Scale bar, $50\;\mu$m.
}
	\label{SMovie:lattice}
\end{figure}

\begin{figure}[h]
	\centering
	\includegraphics[width=0.5\columnwidth]{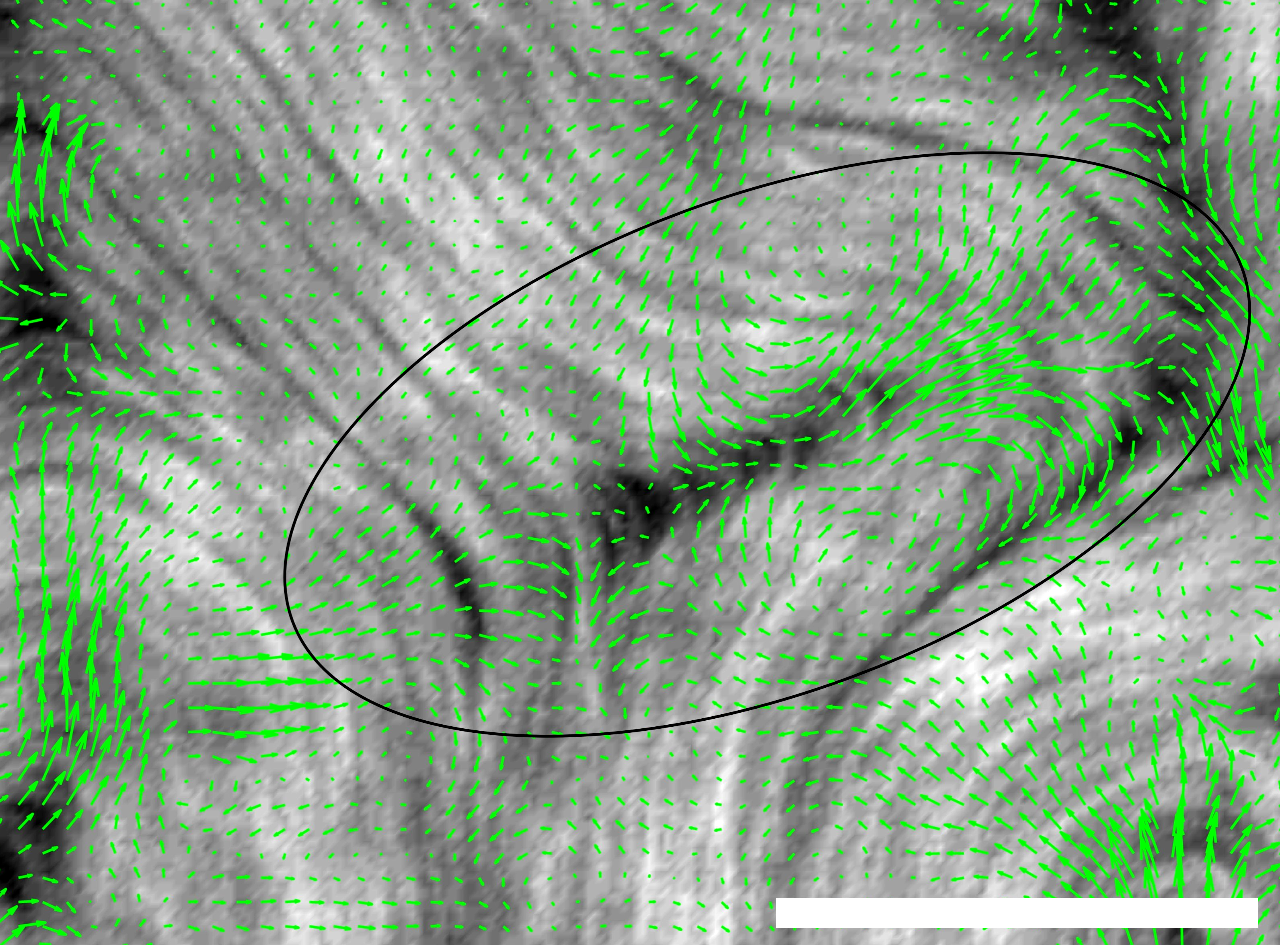}
	\caption{Force density map during the unbinding of a pair of complementary defects. The black ellipse is a guide for the eye to locate the region where the unbinding event takes place. Scale bar is 50 $\mu$m.
}
	\label{SMovie:unbind}
\end{figure}

\begin{figure}[h]
	\centering
	\includegraphics[width=0.5\columnwidth]{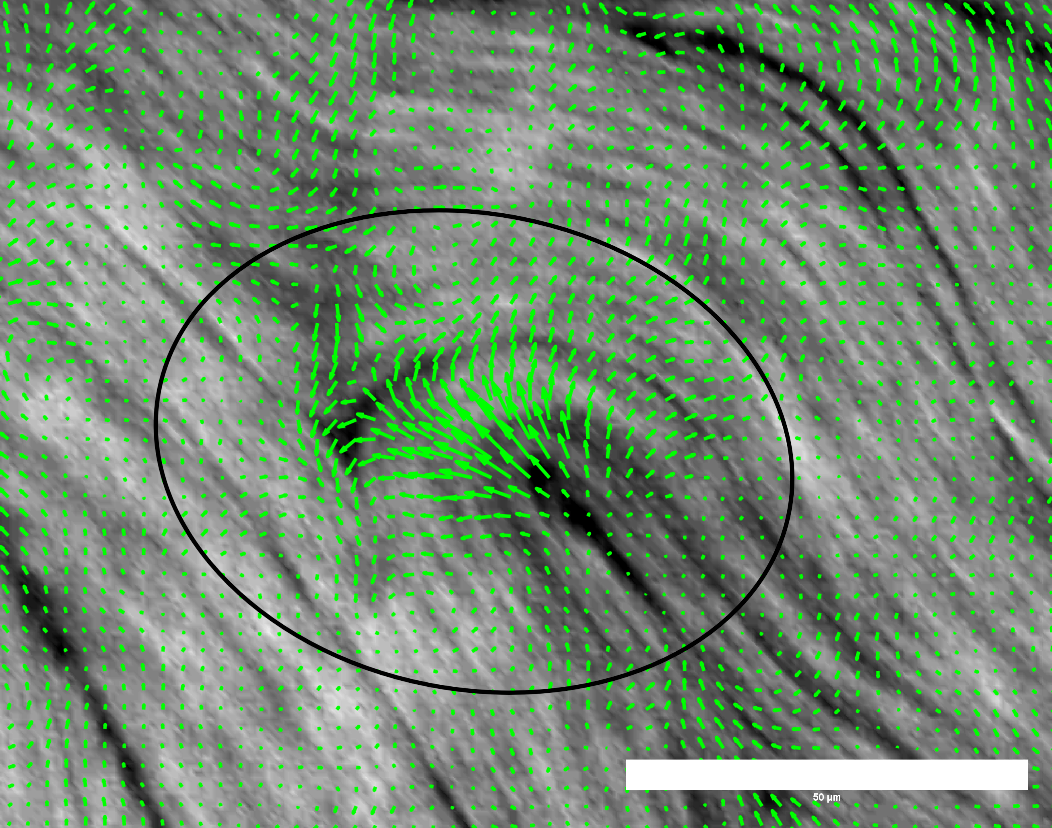}
	\caption{Force density map during the annihilation of a pair of complementary defects. The black ellipse is a guide for the eye to locate the region where the event takes place. Scale bar is 50 $\mu$m.
}
	\label{SMovie:bind}
\end{figure}

\begin{figure}[h]
	\centering
	\includegraphics[width=0.25\columnwidth]{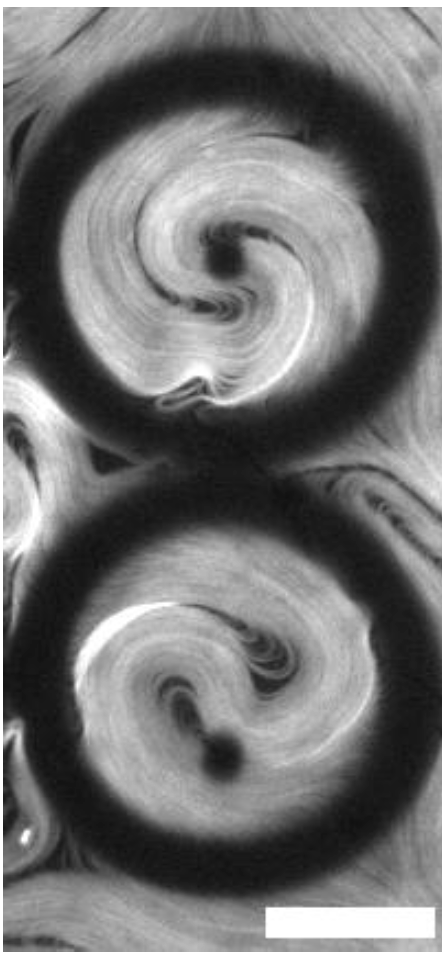}
	\caption{Hybrid microfabrication of a compliant column surrounded by a rigid ring. In both cases the ring confines a $+1/2$ defect. Top: centered column. Bottom: off-centered column. Scale bar is 50 $\mu$m.
}
	\label{SMovie:offcentered}
\end{figure}

\begin{figure}[h]
	\centering
	\includegraphics[width=0.4\columnwidth]{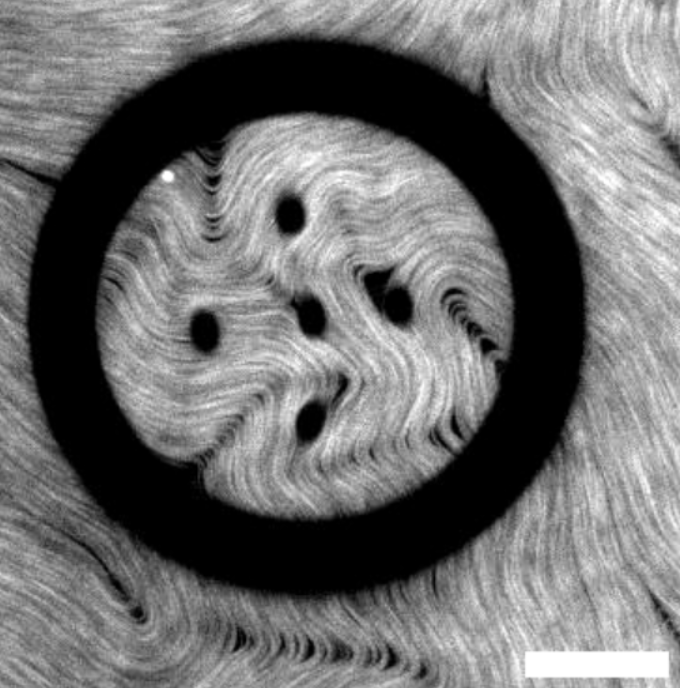}
	\caption{Hybrid microfabrication of a set of compliant columns surrounded by a rigid ring that confines a $-1/2$ defect. Scale bar is 50 $\mu$m.
}
	\label{SMovie:negative}
\end{figure}

\end{document}